\documentclass[aps,prd,amsmath,twocolumn,amssymbaps,showpacs]{revtex4-2}
\usepackage{graphicx}  
\usepackage{dcolumn}   
\usepackage{bm}        
\usepackage{amssymb}   
\usepackage{amsmath}   
\usepackage{bbm}
\usepackage{draftcopy}
\usepackage{relsize} 
\usepackage{xfrac}    
\usepackage{slashed}
\usepackage{color}
\usepackage{footnote}
\definecolor{dgreen}{cmyk}{1.,0.,1.,0.2}        
\definecolor{orange}{cmyk}{0.,0.353,1.,0.}    

\usepackage[bookmarks]{hyperref}

\newcommand{\di}{{\rm d}}

\newcommand{\be}{\begin{equation}}
\newcommand{\ee}{\end{equation}}                                                                               
\newcommand{\bea}{\begin{eqnarray}}
\newcommand{\eea}{\end{eqnarray}} 

\begin{document}
\title{QCD phase transition at finite isospin density and magnetic field within the three-flavor NJL model}

\author{Gaoqing Cao$^{1,2}$}
\affiliation{1 School of Physics and Astronomy, Sun Yat-sen University, Zhuhai 519082, China\\
2 Guangdong Provincial Key Laboratory of Quantum Metrology and Sensing, Sun Yat-Sen University, Zhuhai 519082 
China}
\date{\today}

\begin{abstract}
Previously, the QCD phase transition at finite isospin density and magnetic field was explored within the two-flavor Nambu--Jona-Lasinio model. This work extends the study to the more realistic three-flavor case, where not only strange quark contributions but also quark mass splitting in a strong magnetic field are fully taken into account. Adopting the Ginzburg-Landau approximation and the Landau representation for fermion propagators, we re-explore the transitions from the normal chiral symmetry breaking phase to pion superfluidity or rho superconductivity. Unlike the previous study, we project the mesonic fields onto the eigenstates of a charged point particle in a magnetic field and prove that the corresponding self-energies from quark loops are gauge invariant and degenerate with respect to the extra transverse degrees of freedom. However, the numerical results are qualitatively consistent with previous findings: as the isospin chemical potential increases, pion superfluidity is favored at small magnetic fields, while rho superconductivity is favored at large magnetic fields. In the three-flavor model, since the lowest energy of the rho meson increases with stronger magnetic field, the corresponding critical isospin chemical potential also increases with the magnetic field.
\end{abstract}


\maketitle

\section{Introduction}

A central goal of modern nuclear physics is to employ Quantum Chromodynamics (QCD) and its effective theories to map the rich phase structure of nuclear matter across the parameter space spanned by temperature, baryon density, and other variables \cite{Klevansky:1992qe,Stephanov:2004wx, Fukushima:2010bq,Luo:2017faz,Son:2000xc,Kogut:2002zg,He:2005nk,Chen:2015hfc,Jiang:2016wvv,Cao:2021rwx}. In particular, strong magnetic fields of up to $10^{18}-10^{20} {\rm G}$ can be generated in the early Universe, magnetars~\cite{Bocquet:1995je}, and non-central collisions \cite{Skokov:2009qp,Deng:2012pc,STAR:2023jdd, Huang:2024hua, Zhao:2022dac, Shen:2025unr}, where finite isospin density may also play a role. One key feature of QCD is chiral symmetry breaking and its restoration, and it is well established that magnetic fields strengthen the chiral condensate at low temperatures, known as the ``magnetic catalysis effect'' \cite{Gusynin:1994xp}. However, first-principles lattice QCD simulations reveal that magnetic fields suppress the chiral condensate near the crossover temperatures, known as the ``inverse magnetic catalysis effect'' \cite{Bali:2011qj, Bali:2012zg}. Furthermore, the spectra of charged hadrons are significantly affected by strong magnetic fields~\cite{Bali:2017ian,Ding:2020hxw,Ding:2026qzu}, thus instabilities can arise under mixed conditions, calling for more precise determinations of the ground-state phases~\cite{Cao:2021rwx}.

At a finite isospin chemical potential $\mu_{\rm I}$, it has been well established that the QCD vacuum undergoes a transition to charged pion $(\pi^\pm)$ superfluidity for $\mu_{\rm I} > m_\pi$ \cite{Son:2000xc, Kogut:2002zg,He:2005nk,Brandt:2017oyy}, within which a BEC-BCS crossover occurs at larger $\mu_{\rm I}$~\cite{Sun:2007fc,Cuteri:2021hiq}. While charged rho mesons $\rho^\pm$ share the same isospin quantum numbers as $\pi^\pm$, $\rho^\pm$ superconductivity is suppressed by $\pi^\pm$ superfluidity for however large $\mu_{\rm I}$~\cite{Brauner:2016lkh}. Nevertheless, the situation could be altered when a strong magnetic field is present together with a large $\mu_{\rm I}$. Both lattice QCD and effective theory studies~\cite{Bali:2017ian,Cao:2019res,Cao:2021rwx,Liu:2026kvs} have found that the lowest energy of $\pi^\pm$ exceeds that of $\rho^\pm$ at large magnetic field, thus the critical $\mu_{\rm I}$ for $\rho^\pm$ superconductivity will be smaller than that for $\pi^\pm$ superfluidity according to the silver-blaze properties~\cite{Cohen:2003kd}. Consequently, $\rho^\pm$ superconductivity will become the favored ground state for large magnetic field and large $\mu_{\rm I}$~\cite{Ke:2026npb}. As references, such an inversion has been found in two relevant systems: one with parallel rotation and magnetic field~\cite{Liu:2017spl,Cao:2019ctl,Chen:2019tcp,Cao:2020pmm}, and the other with rotation and finite isospin chemical potential~\cite{Zhang:2018ome, Zhang:2020drr}.
 
As has been discussed in detail in our previous work~\cite{Ke:2026npb}, $\pi^\pm$ superfluidity and $\rho^\pm$ superconductivity are type-II superconducting phases in an external magnetic field, thus we are justified to employ the Ginzburg-Landau (GL) approximation to calculate the phase boundaries between these phases and the normal chiral symmetry breaking phase. This work extends the previous one to the more realistic three-flavor case and more physically adopts the Landau-eigenstate projecting scheme~\cite{Liu:2026kvs} to calculate the self-energies of $\pi^\pm$ and $\rho^\pm$. This paper is organized as follows. In Sec.~\ref{NJL}, the whole formalism is developed with the gap equations and mesonic properties analytically explored in Secs.~\ref{gap} and~\ref{meson}, respectively. Within Sec.~\ref{meson}, the inverse propagators of the lowest-Landau-level mesons are firstly derived in Sec.~\ref{prop}, and then the corresponding Ginzburg-Landau coefficients are regularized in detail in Sec.~\ref{GLsL}. Finally, numerical results are presented in Sec.~\ref{num} and we summarize in Sec.~\ref{summary}. Additionally, two appendices are attached at the end: In Appendix~\ref{degen}, we prove that the self-energies of mesons are degenerate with respect to the extra transverse degrees of freedom, such as the longitudinal angular momentum in the symmetric gauge. In Appendix~\ref{prop2}, we derive the explicit forms of the inverse propagators of mesons based on the Landau representation for fermion propagators, and give some examples for several Landau levels of mesons.

\section{Three-flavor Nambu--Jona-Lasinio model}\label{NJL}

\subsection{Thermodynamic potential and gap equations}\label{gap}

In the three-flavor Nambu--Jona-Lasinio (NJL) model, the mass splitting between $u$ and $d$ quarks in a magnetic field avoids the occurrence of vacuum superconductivity~\cite{Cao:2019res}, which is consistent with lattice QCD simulations~\cite{Bali:2017ian,Hidaka:2012mz}. Thus, the three-flavor NJL model is more suitable for realistic studies, especially when issues relevant to vector mesons are involved. By extending the original Lagrangian density~\cite{Klevansky:1992qe,Klimt:1989pm} with additional four-fermion (axial-)vector interaction terms, the full Lagrangian in Euclidean space follows as
\begin{eqnarray}
{\cal L}_{\rm NJL}&=&\bar\psi\left(i\slashed{D}-m_0+\frac{\mu _I}{2}\gamma^0\lambda _3 \right)\psi+G_S\left[\left(\bar\psi\lambda^a\psi\right)^2+\right.\nonumber\\
&&\left.\left(\bar\psi i\gamma_5\lambda^a\psi\right)^2\right]\!-\!G_V\!\left[\left(\bar\psi\gamma^\mu\lambda^a\psi\right)^2\!+\!\left(\bar\psi i\gamma^\mu\gamma_5\lambda^a\psi\right)^2\right]\nonumber\\
&&+\ {\cal L}_6,\ \ \ \ {\cal L}_6=-K\sum_{s=\pm}{\rm Det}\,\left(\bar\psi_i\Gamma^s\psi_j\right),\label{LNJL}
\end{eqnarray} 
where summations over the flavor index $a=(0,\dots,8)$ and Lorentz index $\mu=(1,\dots,4)$ should be understood. Here, $\psi=(u,d,s)^T$ represents the three-flavor quark field, $m_0={\rm diag}(m_{\rm 0u},m_{\rm 0d},m_{\rm 0s})$ is the current quark mass matrix, and the covariant derivative is defined as $D_\mu=\partial_\mu-iQA_\mu$ with the charge matrix $Q={\rm diag}(q_{\rm u},q_{\rm d},q_{\rm s})$. In flavor space, $\lambda^0=\sqrt{2\over3}I$ and $\lambda^i~(i=1,\dots,8)$ are Gell-Mann matrices. The six-fermion interaction term ${\cal L}_6$ is introduced to account for the $U_A(1)$ anomaly effect on the (pseudo-)scalar channels~\cite{tHooft:1976snw}, where $\Gamma^\pm=1\pm\gamma_5$ and the determinant is applied to the matrix with elements $\bar\psi_i\Gamma^s\psi_j\ (i,j=1,2,3)$. Note that the number index $i=1,2,3$ corresponds to the Latin index ${ f=u,d,s}$ in flavor space.

Now, following the Ginzburg-Landau approximation as adopted in the two-flavor case, we only consider the nonzero chiral condensations $\sigma_{\rm i}\equiv\langle\bar{\psi}^i{\psi}^i\rangle$, and then calculate the GL coefficients based on their expectation values. For this purpose, it is convenient to reduce ${\cal L}_6$ to an effective four-fermion interaction form under the Hartree approximation~\cite{Klevansky:1992qe}, as
\begin{widetext}
\bea
{\cal L}_6^4&=&-{K\over6}\Big\{2\sum_{\rm f=u,d,s}\sigma_f(\bar{\psi}\lambda^0\psi)^2-3\sigma_s\sum_{i=1}^3(\bar{\psi}\lambda^i\psi)^2
-3\sigma_{\rm d}\sum_{i=4}^5(\bar{\psi}\lambda^i\psi)^2-3\sigma_{\rm u}\sum_{i=6}^7(\bar{\psi}\lambda^i\psi)^2+(\sigma_s\!-\!2\sigma_{\rm u}\!-\!2\sigma_{\rm d})(\bar{\psi}\lambda^8\psi)^2\nonumber\\
&&+\sqrt{2}(2\sigma_s\!-\!\sigma_{\rm u}\!-\!\sigma_{\rm d})(\bar{\psi}\lambda^0\psi)(\bar{\psi}\lambda^8\psi)-\sqrt{6}(\sigma_{\rm u}\!-\!\sigma_{\rm d})(\bar{\psi}\lambda^3\psi)(\bar{\psi}\lambda^0\psi-\sqrt{2}\bar{\psi}\lambda^8\psi)\Big\}-(\lambda^a\rightarrow i\lambda^a\gamma^5).
\eea
And the three-flavor Lagrangian becomes effectively one with only four-fermion effective interactions, just like that of the two-flavor NJL model,
\begin{eqnarray}
{\cal L}_{\rm NJL}^4\!=\!\bar\psi(i\slashed{D}\!-\!m_0\!+\!\frac{\mu _I}{2}\gamma^0\lambda _3)\psi\!+\!\!\left[G_{ab}^-\left(\bar\psi\lambda^a\psi\right)\left(\bar\psi\lambda^b\psi\right)\!\!+\!G_{ab}^+\left(\bar\psi i\gamma_5\lambda^a\psi\right)\left(\bar\psi i\gamma_5\lambda^b\psi\right)\right]\!\!-\!G_V\!\!\left[\left(\bar\psi\gamma^\mu\lambda^a\psi\right)^2\!+\!\left(\bar\psi i\gamma^\mu\gamma_5\lambda^a\psi\right)^2\right],\nonumber\\\label{NJL4}
\end{eqnarray}
where the non-vanishing elements of the symmetric coupling matrices $G^\pm$ are given by~\cite{Klevansky:1992qe,Cao:2019res}
\begin{eqnarray}
&&G_{00}^\mp=G_S\mp {K\over3}\sum_{\rm f=u,d,s}\sigma_f,~G_{11}^\mp=G_{22}^\mp=G_{33}^\mp=G_S\pm {K\over2}\sigma_s,~G_{44}^\mp=G_{55}^\mp=G_S\pm {K\over2}\sigma_{\rm d},~G_{66}^\mp=G_{77}^\mp=G_S\pm {K\over2}\sigma_{\rm u},\nonumber\\
&&G_{88}^\mp=G_S\mp {K\over6}(\sigma_s-2\sigma_{\rm u}-2\sigma_{\rm d}),~G_{08}^\mp=\mp {\sqrt{2}K\over12}(2\sigma_s\!-\!\sigma_{\rm u}\!-\!\sigma_{\rm d}),~G_{38}^\mp=-\sqrt{2}G_{03}^\mp=\mp {\sqrt{3}K\over6}(\sigma_{\rm u}\!-\!\sigma_{\rm d}).
\end{eqnarray}

By following the derivations in Ref.~\cite{Cao:2019res} and adopting the so-called ``vacuum regularization'', the thermodynamic potential can be given by
\begin{eqnarray}
	\Omega&=&2G_S\sum_{{\rm f}=u,d,s}\sigma_f^2-4K\prod_{{\rm f}=u,d,s}\sigma_f-N_c\sum_{{\rm f}=u,d,s}\left\{{{m_f}^4\over8\pi^2}\Big[\tilde{m}_f^{-1}\Big(1+{2\tilde{m}_f^{-2}}\Big)\Big({1+{\tilde{m}_f^{-2}}}\Big)^{1\over2}-\ln\Big({\tilde{m}_f^{-1}}
+\Big({1+{\tilde{m}_f^{-2}}}\Big)^{1\over2}\Big)\Big]\right.\nonumber\\
&&\left.-{1\over8\pi^2}\int_0^\infty {ds\over s^3}e^{-{m_f}^2s}\left({q_fBs
	\over\tanh(q_fBs)}-1\right)\right\}
\end{eqnarray}
with the dynamical quark masses $m_i=m_{0i}-4G_S\sigma_i+K\sum_{jk}\!\epsilon_{ijk}^2\sigma_j\sigma_k$ and the corresponding reduced masses defined by $\tilde{m}_f\equiv{m_f/\Lambda}$. Then, the gap equations can be derived self-consistently from the extremal conditions $\partial\Omega/\partial \sigma_f=0$ as
\begin{eqnarray}
-\sigma_f
&=&N_c{{m_f}^3\over2\pi^2}\Big[\tilde{m}_f^{-1}\Big({1+\tilde{m}_f^{-2}}\Big)^{1\over2}-\ln\Big({\tilde m}_f^{-1}
+\Big({1+\tilde{m}_f^{-2}}\Big)^{1\over2}\Big)\Big]+N_c{m_f\over4\pi^2}\int_0^\infty {ds\over s^2}e^{-{m_f}^2s}\left({q_fBs
	\over\tanh(q_fBs)}-1\right).\label{mgap}
\end{eqnarray}
\end{widetext}
\subsection{Mesons in the random phase approximation}\label{meson}

Based on the effective Lagrangian \eqref{NJL4}, the mesonic properties can be studied in the standard random phase approximation (RPA)~\cite{Klevansky:1992qe}. By taking the Hubbard-Stratonovich transformation with the help of auxiliary fields,
\begin{equation}
\begin{aligned}
S^a &\equiv-2G_{\rm aa}^-\,\bar{\psi}\lambda^a\psi, &
P^a &\equiv-2G_{\rm aa}^+\,\bar{\psi}i\gamma_5\lambda^a\psi,\\
V^{a\mu} &\equiv-2G_{\rm V}\,\bar{\psi}\gamma^{\mu}\lambda^a\psi, &
A^{a\mu} &\equiv-2G_{\rm V}\,\bar{\psi}i\gamma^{\mu}\gamma_5\lambda^a\psi,
\end{aligned}
\end{equation}
the Lagrangian \eqref{NJL4} can be equivalently transformed to the form
\bea
\mathcal{L}&=& \bar{\psi}\Bigl( i\slashed{D}\!-\!m_0\!+\!\frac{\mu_{\rm I}}{2}\gamma^0\lambda^3\!\!-\!\left(S^a \!\!+\!i\gamma_5P^a\!\!-\!\slashed{V}^{a}\!\!-\!i\gamma_5\slashed{A}^{a}\right)\lambda^a\Bigr)\psi \nonumber\\
&&-\frac{(S^a)^2}{4G_{aa}^-}-\frac{(P^a)^2}{4G_{aa}^+}+\frac{V^{a\mu}V_{\mu}^{ a}+A^{a\mu}A_{\mu}^{ a}}{4G_{\rm V}}+\cdots
\eea
with $\slashed{X}\equiv X^\mu\gamma_\mu$ and ``$\cdots$'' denoting the mixings among the neutral (pseudo-)scalar fields in the channels $a=0,3,8$. In the presence of a magnetic field, it is more convenient to present the mesonic fields in electric-charge eigenstates, that is, we redefine the pseudoscalar and vector fields as
\begin{widetext}
\bea
&&\eta^0=P^0, \pi^0 =P^3, \eta^8=P^8, \pi^{\pm} \!=\!\frac{1}{\sqrt{2}}(P^1\!\mp\! iP^2), K^\pm\!=\!\frac{1}{\sqrt{2}}(P^4\!\mp\! iP^5), K^0\!=\!\frac{1}{\sqrt{2}}(P^6\!+\!iP^7), \bar{K}^0\!=\!\frac{1}{\sqrt{2}}(P^6\!-\!iP^7)\nonumber\\
&&\omega =V^{ 0}, \rho^{0} =V^{3}, \phi_8=V^{ 8},
\rho^{\pm}\!=\!\frac{1}{\sqrt{2}}(V^{1}\!\mp\! iV^{2}), K^{*\pm}\!=\!\frac{1}{\sqrt{2}}(V^{4}\!\mp\! iV^{5}), K^{*0}\!=\!\frac{1}{\sqrt{2}}(V^{6}\!+\!iV^{7}), \bar{K}^{*0}\!=\!\frac{1}{\sqrt{2}}(V^{6}\!-\!iV^{7}).
\eea
Then, the Lagrangian can be rewritten as
\bea
\mathcal{L}&=&\bar{\psi}\Bigl[ i\slashed{D}\!-\!m_0\!+\!\frac{\mu_{\rm I}}{2}\gamma^0\lambda^3\!-\!S^a\lambda^a \!-\!i\gamma_5(\eta^0\lambda^0 +\pi^0\lambda^3 +\eta^8\lambda^8+\pi^\pm\lambda_{12}^\pm+K^\pm\lambda_{45}^\pm+K^0\lambda_{67}^++ \bar{K}^0\lambda_{67}^-)\nonumber\\
&&+\slashed{\omega}\lambda^0+\slashed{\rho}^0\lambda^3+\slashed{\phi}_8\lambda^8+\slashed{\rho}^\pm\lambda_{12}^\pm+\slashed{K}^{*\pm}\lambda_{45}^\pm+\slashed{K}^{*0}\lambda_{67}^++\bar{\slashed{K}}^{*0}\lambda_{67}^-+\slashed{A}^{a}\!\lambda^a i\gamma_5\Bigr]\psi \nonumber\\
&&-\frac{(S^a)^2}{4G_{aa}^-}-\frac{(\eta^0)^2}{4G_{00}^+}-\frac{(\pi^0)^2}{4G_{33}^+}-\frac{(\eta^8)^2}{4G_{88}^+}-\frac{\pi^+\pi^-}{2G_{11}^+}-\frac{K^+K^-}{2G_{44}^+}-\frac{\bar{K}^0K^0}{2G_{66}^+}\nonumber\\
&&+\frac{1}{4G_{\rm V}}\left[(\omega^\mu)^2+(\rho^{0\mu})^2+(\phi_8^\mu)^2+2\rho^{+\mu}\rho^{-}_{\mu}+2K^{*+\mu}K^{*-}_{\mu}+2\bar{K}^{*0\mu}K^{*0}_{\mu}+A^{a\mu}A_{\mu}^{ a}\right]+\cdots
\eea
\end{widetext}
with $\lambda_{ij}^\pm  \equiv (\lambda_i\pm i\lambda_j)/\sqrt{2}$. In the mean field approximation with nonzero chiral condensates $\sigma_i$, only the following part of the Lagrangian
\bea
\!\!\!\!\!\!\!\mathcal{L}'&=&\bar{\psi}\Bigl[ {\cal G}^{-1}\!-\!i\gamma_5\pi^\pm\lambda_{12}^\pm\!+\!\slashed{\rho}^\pm\lambda_{12}^\pm\Bigr]\psi\!-\!\frac{\pi^+\pi^-}{2G_{11}^+}\!+\!\frac{\rho^{+\mu}\rho^{-}_{\mu}}{2G_{\rm V}}\label{Lagp}
\eea
is relevant to our study of pion superfluidity and rho superconductivity. Here, ${\cal G}^{-1}\equiv i\slashed{D}-m+\frac{\mu_{\rm I}}{2}\gamma^0\lambda^3$ is the inverse quark propagator in flavor space, that is, ${\cal G}={\rm diag}(G_{u},G_{d}, G_{s})$, and $m={\rm diag}(m_{\rm u},m_{\rm d},m_{\rm s})$ is the corresponding dynamical mass matrix. By completing the integrations over all fermion fields, the corresponding action can be bosonized as
\bea
\mathcal{S}'&=&i\ {\rm Tr} \ln\Bigl[ {\cal G}^{-1}-\!i\gamma_5\pi^\pm\lambda_{12}^\pm+\slashed{\rho}^\pm\lambda_{12}^\pm\Bigr]\nonumber\\
&&\ \ \ \ \ \ \ \ \ \ \ \ \ \ \ +\int \di^4 x\left(\frac{\pi^+\pi^-}{2G_{11}^+}-\frac{\rho^{+\mu}\rho^{-}_{\mu}}{2G_{\rm V}}\right),
\eea
where the trace ${\rm Tr}$ is over coordinate, Dirac, flavor, and color spaces.

In the standard random phase approximation, we expand $\pi^\pm(x)$ over the energy-momentum basis $e^{-i\,q\cdot x}$ for an isotropic system, and then evaluate their dispersion relations by requiring the effective propagator to be divergent. When an external magnetic field is applied, the treatment becomes subtle, as gauge invariance must be strictly preserved in the investigation of their physical properties. Previously, we took care of the gauge invariance of the self-energy by simply compensating a Schwinger phase from $\pi^\pm(x)$ to the quark loops~\cite{Cao:2015xja}. Though the well-known results in the vanishing $eB$ limit can be well reproduced, such a treatment is subject to ambiguity when both a magnetic field and rotation are present in the system, where the Schwinger phase cannot be separately identified from the quark loops~\cite{Cao:2019ctl,Chen:2019tcp}. However, if we expand $\pi^\pm(x)$ over the Landau basis with well-defined longitudinal energy-momentum and Landau levels~\cite{Liu:2026kvs}, the gauge invariance is automatically guaranteed since each interaction vertex is totally neutral. Take the interaction term $$\bar{\psi}(x)i\gamma_5\pi^+(x)\lambda_{12}^+\psi(x)=\sqrt{2}\,\pi^+(x)\bar{u}(x)i\gamma_5d(x)$$ in \eqref{Lagp} for example, the corresponding term on the Landau basis is $\pi^+_n(x)\bar{u}_{n_1}(x)i\gamma_5d_{n_2}(x)$ with $n, n_1, n_2$ Landau levels. Under any gauge transformation, $A_\mu\rightarrow A_\mu+\Delta A_\mu$, the corresponding eigenstates change as
\bea
\pi^+_n(x)&\rightarrow& e^{-ie\int_{x_0}^x\Delta A\cdot \di z}\pi^+_n(x),\nonumber\\
\bar{u}_{n_1}(x)&\rightarrow& e^{iq_{u}\int_{x_0}^x\Delta A\cdot \di z}  \bar{u}_{n_1}(x), \nonumber\\
d_{n_2}(x)&\rightarrow& e^{-iq_{d}\int_{x_0}^x\Delta A\cdot \di z} {d}_{n_2}(x),
\eea
thus the leading Schwinger phases exactly cancel out in the interaction term. For the more complicated case with both a magnetic field and rotation, the leading Schwinger phases must be defined in curved space but the gauge invariance remains. The same holds for $\rho^\pm(x)$, but the Lorentz indices $\mu$ must be rearranged based on the spin eigenstates in order to get rid of the mixings introduced by the magnetic field. It follows that the relevant action can be reduced to
\bea
\mathcal{S}'\!&=&\!i\, {\rm Tr} \ln\!\Big[ {\cal G}^{-1}\!\!+\!\Big(\!\!-\!i\gamma_5\pi^\pm\!+\!{\rho}^\pm_0\gamma^0\!\!+\!{\rho}^\pm_3\gamma^3\!\!+\!\!\!\sum_{s=\pm 1}\!{\rho}^\pm_{s}\gamma^{s}_{12}\Big)\lambda_{12}^\pm\Big]\nonumber\\
&&\!\!+\!\int\!\! \di^4 x\!\left(\frac{\pi^+\pi^-}{2G_{11}^+}\!-\!\frac{\rho^{+}_{0}\rho^{-}_{0}\!-\!\rho^{+}_{3}\rho^{-}_{3}\!-\!\sum_{s=\pm 1}\rho^{+}_{s}\rho^{-}_{-s}}{2G_{\rm V}}\right)
\eea
with $\gamma^{s}_{12}\equiv {\gamma^1+ s\,i\,\gamma^2\over\sqrt{2}}$ and the index $s=\pm 1$ the spins along the direction of the magnetic field.

\begin{widetext}
\subsubsection{Inverse propagators of $\pi^\pm$ and $\rho^{\pm}_{\pm1}$}\label{prop}
We will focus on the most relevant modes, $\pi^\pm$ and $\rho^{\pm}_{\pm1}$; then Taylor expansions over these fields give the quadratic terms as
 \bea
\!\!\mathcal{S}_2'&=&\int\! \di^4 x\left(\frac{\pi^+\pi^-}{2G_{11}^+}\!+\!\frac{\rho^{+}_{1}\rho^{-}_{-1}}{2G_{\rm V}}\right)-i\, {\rm Tr} \big({\cal G}\,i\gamma_5\pi^+\lambda_{12}^+\,{\cal G}\,i\gamma_5\pi^-\lambda_{12}^-\big)-i\, {\rm Tr} \big({\cal G}\,{\rho}^+_{1}\gamma^{+1}_{12}\lambda_{12}^+\,{\cal G}\,{\rho}^-_{-1}\gamma^{-1}_{12}\lambda_{12}^-\big)\nonumber\\
&=&\int\! \di^4 x\left(\frac{\pi^+(x)\pi^-(x)}{2G_{11}^+}\!+\!\frac{\rho^{+}_{1}(x)\rho^{-}_{-1}(x)}{2G_{\rm V}}\right)-2 N_c\int\di^4 x \int\di^4 x'\ \pi^+(x')\ {\rm tr}\big[{G}_u(x,x')\,i\gamma_5\,{G}_d(x',x)\,i\gamma_5\big]\ \pi^-(x)\nonumber\\
&&-2N_c\int\di^4 x \int\di^4 x' \ {\rho}^+_{1}(x')\  {\rm tr}\big[{G}_u(x,x')\,\gamma^{+1}_{12}\,{G}_d(x',x)\,\gamma^{-1}_{12}\big]\ {\rho}^-_{-1}(x), 
\eea
where the trace ${\rm tr}$ is now over coordinate and Dirac spaces. From now on, we choose the symmetric gauge with the vector potential $A_\mu=(0,B\,y/2,-B\,x/2,0)$; then both the longitudinal $z$-direction angular momentum $l$ and Landau level $n$ are well-defined quantum numbers. The corresponding eigenstates of $\pi^+(x)$ and $\rho^{+}_{1}(x)$ can be given as $M_{n,l}(q_\parallel,x)\equiv e^{-i\,q_\parallel\cdot x_\parallel}\chi_n^{l}(eB,x_\bot)$, where $q_\parallel=(q_3,q_4)$ is the longitudinal energy-momentum, $q_\parallel\cdot x_\parallel=-q_3x_3-q_4x_4,$ and $\chi_n^{l}(eB,x_\bot)\equiv \left[{eB\over2\pi}{ n!\over(n+l)!}\right]^{1\over2}{e^{i\, l\theta}}\left({eB{r}^2\over2}\right)^{l\over2}e^{-{eB{r}^2\over4}}L_n^l\left({eB{r}^2\over2}\right)$ are the eigenstates for transverse dynamics~\cite{Chen:2015hfc,Liu:2017spl,Cao:2019ctl}. Then, the quadratic terms can be rewritten in longitudinal energy-momentum and Landau spaces as
 \bea
\!\!\mathcal{S}_2'
&=&{1\over V_\bot}\sum_{n=0}^\infty\sum_{l=-n}^\infty\int{\di^2q_\parallel\over(2\pi)^2}\left[\hat{\pi}^+_{n,l}(q_\parallel)D^{-1}_{\pi^\pm}(q_\parallel,n,l)\hat{\pi}^-_{n,l}(q_\parallel)+\hat{\rho}^{+}_{1, (n,l)}(q_\parallel)D^{-1}_{\rho^\pm_{\pm1}}(q_\parallel,n,l)\hat{\rho}^{-}_{-1, (n,l)}(q_\parallel)\right]
\eea
with $V_\bot$ the transverse area and the inverse propagators
 \bea
\!\!\!D^{-1}_{\pi^\pm}&=&{1\over2G_{11}^+}-2N_c\!\int\!\di^4 x \!\int\!\di^4 x'\ {e^{-i[q_3(x_3-x_3')+q_4(x_4-x_4')]}}\chi_n^{l*}(eB,x_\bot)\chi_n^{l}(eB,x_\bot'){\rm tr}\big[{G}_u(x,x')\,i\gamma_5\,{G}_d(x',x)\,i\gamma_5\big],\label{Dpi0}\\
\!\!\!D^{-1}_{\rho^\pm_{\pm1}}&=&{1\over2G_V}-2N_c\!\int\!\di^4 x \!\int\!\di^4 x'\ {e^{-i[q_3(x_3-x_3')+q_4(x_4-x_4')]}}\chi_n^{l*}(eB,x_\bot)\chi_n^{l}(eB,x_\bot')\ {\rm tr}\big[{G}_u(x,x')\,\gamma^{+1}_{12}\,{G}_d(x',x)\,\gamma^{-1}_{12}\big].\label{Drho0}
\eea

In Appendix~\ref{degen}, we prove that $D^{-1}_{\pi^\pm}(q_\parallel,n,l)$ and $D^{-1}_{\rho^\pm_{\pm1}}(q_\parallel,n,l)$ are $l$ independent, consistent with the point-particle results. Thus, by redefining the mesonic fields by their averages, $\hat{\pi}^+_{n}\hat{\pi}^-_{n}\equiv \sum_{l=-n}^\infty\hat{\pi}^+_{n,l}\hat{\pi}^-_{n,l}/(\sum_{l=-n}^\infty)$ and $\hat{\rho}^{+}_{1,n}\hat{\rho}^{-}_{-1,n}\equiv \sum_{l=-n}^\infty\hat{\rho}^{+}_{1, (n,l)}\hat{\rho}^{-}_{-1, (n,l)}/(\sum_{l=-n}^\infty)$, the quadratic terms can be further reduced to 
 \bea
\!\!\mathcal{S}_2'
&=&{eB\over2\pi}\sum_{n=0}^\infty\int{\di^2q_\parallel\over(2\pi)^2}\left[\hat{\pi}^+_{n}(q_\parallel)D^{-1}_{\pi^\pm}(q_\parallel,n)\hat{\pi}^-_{n}(q_\parallel)+\hat{\rho}^{+}_{1,n}(q_\parallel)D^{-1}_{\rho^\pm_{\pm1}}(q_\parallel,n)\hat{\rho}^{-}_{-1,n}(q_\parallel)\right]
\eea
after summing over $l$. Here, the inverse propagators are modified to
\bea
\!\!\!D^{-1}_{\pi^\pm}&=&{1\over2G_{11}^+}-2N_c\int{\di^4\Delta x}\ {e^{-i(q_3\Delta x_3+q_4\Delta x_4)}}P_{n}({\bf \Delta x}_\bot)\ {\rm tr}\big[\tilde{G}_u(\Delta x)\,i\gamma_5\,\tilde{G}_d(-\Delta x)\,i\gamma_5\big],\label{Dpi}\\
\!\!\!D^{-1}_{\rho^\pm_{\pm1}}&=&{1\over2G_V}-2N_c\int{\di^4\Delta x}\ {e^{-i(q_3\Delta x_3+q_4\Delta x_4)}}P_{n}({\bf \Delta x}_\bot)\  {\rm tr}\big[\tilde{G}_u(\Delta x)\,\gamma^{+1}_{12}\,\tilde{G}_d(-\Delta x)\,\gamma^{-1}_{12}\big],\label{Drho}
\eea
where the transversal projecting function is defined as $P_{n}({\bf \Delta x}_\bot)\equiv e^{-{eB\over4} {\bf \Delta x}_\bot^2} L_n\left({eB\over2}{\bf \Delta x}_\bot^2\right)=P_{n}(-{\bf \Delta x}_\bot)$ with $\Delta x\equiv x-x'$. The effective quark propagators $\tilde{G}_f(\Delta x)$ are related to the full propagators as $G_{f}(x,x')=e^{ i\Phi(q_fB,x_\bot,x'_\bot)}\tilde{G}_f(\Delta x)$ with $e^{ i\Phi(q_fB,x_\bot,x'_\bot)}=e^{-iq_{\rm f}\int_{x'}^{x} [A_\mu +{1\over2}F_{\mu\nu}(z-x')^\nu]dz^\mu}$ the well-known gauge-dependent Schwinger phase~\cite{Schwinger:1951nm}. As mentioned in our previous work~\cite{Ke:2026npb}, it is more convenient to work with the Landau-level representation\cite{Chodos:1990vv,Miransky:2015ava} in the case with finite isospin chemical potential, then the effective quark propagators take the forms
\bea
	\tilde{G}_f(\Delta x)&=&-{i|q_fB|\over4\pi}\int{\di^2 k_\parallel\over(2\pi)^2}{e^{i[k_3(x_3-x_3')+k_4(x_4-x_4')]}}e^{ -{\left| q_fB \right|\over4} {\bf \Delta x}_\bot^2} \sum_{n=0}^{\infty}{\frac{D_n(q_fB,{\bf \Delta x}_\bot)}{\left( k_{4}^{f} \right) ^2+k_{3}^{2}+m^2+2\left| q_fB \right|n}},\nonumber\\
	D_n(q_fB,{\bf \Delta x}_\bot)&=&\left( m-k_{4}^{f}\gamma _4-k_3\gamma _3 \right) \left[ \left( 1+i\gamma _1\gamma _2{S}_{q_fB}\right) L_n\left( {\left| q_fB \right|{\bf \Delta x}_\bot^2\over2} \right) -\left( 1-i\gamma _1\gamma _2{S}_{q_fB} \right) L_{n-1}\left( {\left| q_fB \right|{\bf \Delta x}_\bot^2\over2}\right) \right]\nonumber\\
	&&-2i|q_fB|\left( {\Delta x}_1\gamma _1+{\Delta x}_2\gamma _2 \right) L_{n-1}^{1}\left( {\left| q_fB \right|{\bf \Delta x}_\bot^2\over2}\right)\label{Sfx}
\eea
with $k_{4}^{u/d}=k_{4}\pm i{\mu_{I}\over2}$.

For a given Landau level $n$, the corresponding inverse propagators are $D^{-1}_{\pi^\pm}(q_\parallel,n)$ and $D^{-1}_{\rho^\pm_{\pm1}}(q_\parallel,n)$; thus the property of each Landau level can be investigated term by term in principle, see Appendix~\ref{prop2}. To our concern, the lowest energies are most relevant to the explorations of pion superfluidity and rho superconductivity, so we will focus on the lowest Landau levels of $\pi^\pm$ and $\rho^{\pm}_{\pm1}$ in the following. The corresponding inverse propagators are explicitly
\bea
D^{-1}_{\pi^\pm}(q_\parallel,0)&=&{1\over2G_{11}^+}+{|q_uB|\over4\pi}{|q_dB|\over4\pi}\int{\di^2k_\parallel\over(2\pi)^2}\sum_{n_u=0}^{\infty}\sum_{n_d=0}^{\infty}{N_{\pi^\pm}(0, n_u, n_d; eB, q_uB, q_dB)\over \left[\left( k_{4}^{u} +q_4\right) ^2+E_{n_u}^2(k_3+q_3)\right]\left[\left( k_{4}^{d} \right) ^2+E_{n_d}^2\right]},\\
D^{-1}_{\rho^\pm_{\pm1}}(q_\parallel,0)&=&{1\over2G_V}+{|q_uB|\over4\pi}{|q_dB|\over4\pi}\int{\di^2k_\parallel\over(2\pi)^2}\sum_{n_u=0}^{\infty}\sum_{n_d=0}^{\infty}{N_{\rho^\pm_{\pm1}}(0, n_u, n_d; eB, q_uB, q_dB)\over \left[\left( k_{4}^{u} +q_4\right) ^2+E_{n_u}^2(k_3+q_3)\right]\left[\left( k_{4}^{d} \right) ^2+E_{n_d}^2\right]},
\eea
where the dispersions are given by $E_{n_f}(k_3)\equiv\sqrt{k_3^2+2n_f|q_fB|+m_f^2}$, and the numerators are
\bea
\!\!\!N_{\pi^+}&=&
-2\big[(k_4^{u}+q_4)k_4^{d}\!+\!(k_3\!+\!q_3)k_3\!+\!m_um_d\big]
\left[
h(0,n_u\!-\!1, n_d)\!+\!h(0,n_u, n_d\!-\!1)\right]
\!-\!4\left| q_uB \right|\left| q_dB \right|h(1,n_u\!-\!1, n_d\!-\!1)\nonumber\\
&=&-\left\{\big[(k_4^{u}+q_4)k_4^{d}\!+\!(k_3\!+\!q_3)k_3\!+\!m_um_d\big]\left({\left| q_uB \right|\over n_d}+{\left| q_dB \right|\over n_u}\right)
\!+\!\left|2 q_uB \right|\left| 2q_dB\right|\right\} {h(1,n_u\!-\!1, n_d\!-\!1)},
\\
\!\!\!N_{\bar{\rho}_1^{+}}&=&-4\big[(k_4^{u}+q_4)k_4^{d}\!+\!(k_3\!+\!q_3)k_3\!+\!m_um_d\big]h(0,n_u, n_d)
\eea
with the auxiliary function given by
\bea
h(\alpha, n_u, n_d)
&=&{16\pi\over |eB|}\left({2\over|eB|}\right)^{\alpha}{({n_u}\!+\!{n_d}\!+\!\alpha)!\over{n_u}!\,{n_d}!}
{\left({1\over3}\right)^{n_u}\left({2\over3}\right)^{n_d}}.
\eea

\subsubsection{The Ginzburg-Landau coefficients for the lowest Landau levels}\label{GLsL}
The Ginzburg-Landau coefficients are related to the inverse propagators as ${\cal A}_{\pi^\pm}=D^{-1}_{\pi^\pm}(0,0)$ and ${\cal A}_{\rho^\pm_{\pm1}}=D^{-1}_{\rho^\pm_{\pm1}}(0,0)$~\cite{Ke:2026npb}, thus their explicit expressions can be evaluated as
\bea
{\cal A}_{\pi^\pm}&=&{1\over2G_{11}^+}\!-\!{|q_uB|\over4\pi}{|q_dB|\over4\pi}\!\int{\di^2k_\parallel\over(2\pi)^2}\!\sum_{n_u=0}^{\infty}\sum_{n_d=0}^{\infty}\!{\big(k_4^{u}k_4^{d}\!+\!k_3^2\!+\!m_um_d\big)\left({\left| n_uq_uB \right|}\!+\!{\left| n_dq_dB \right|}\right)
\!+\!\left|2 n_uq_uB \right|\left| 2n_dq_dB \right|\over \left[\left( k_{4}^{u} \right) ^2+E_{n_u}^2\right]\left[\left( k_{4}^{d} \right) ^2+E_{n_d}^2\right]}\bar{h}(1,n_u, n_d)\nonumber\\
&=&{1\over2G_{11}^+}\!-\!{|q_uB|\over4\pi}{|q_dB|\over4\pi}\!\int\!{\di k_3\over 2\pi}\sum_{t=\pm}\sum_{n_u=0}^{\infty}\sum_{n_d=0}^{\infty}\!{\big( t\, \mu_{\rm I}E_{n_u}\!\!-\!m_u\Delta m_l\big)\left({\left| n_uq_uB \right|}\!+\!{\left| n_dq_dB \right|}\right)
\!-\!\left|2n_u q_uB \right|\left({\left| n_uq_uB \right|}\!-\!{\left| n_dq_dB \right|}\right)\over 4E_{n_u}\left[-\left( tE_{n_u}-\,\mu_{\rm I}\right) ^2+E_{n_d}^2\right]}\nonumber\\
&&\times\bar{h}(1,n_u, n_d)\tanh{E_{n_u}+t\,{\mu_{\rm I}\over2}\over 2T}+(u\leftrightarrow d),
\eea
\bea
\!\!\!\!\!\!\!{\cal A}_{\rho^\pm_{\pm1}}&=&{1\over2G_V}-{|q_uB|\over2\pi}{|q_dB|\over2\pi}\int{\di^2k_\parallel\over(2\pi)^2}\sum_{n_u=0}^{\infty}\sum_{n_d=0}^{\infty}{\big(k_4^{u}k_4^{d}\!+\!k_3^2\!+\!m_um_d\big)h(0,n_u, n_d)\over \left[\left( k_{4}^{u} \right) ^2+E_{n_u}^2\right]\left[\left( k_{4}^{d} \right) ^2+E_{n_d}^2\right]}\nonumber\\
&=&{1\over2G_V}\!-\!{|q_uB|\over2\pi}{|q_dB|\over2\pi}\!\int\!{\di k_3\over 2\pi}\!\sum_{t=\pm}\sum_{n_u=0}^{\infty}\sum_{n_d=0}^{\infty}\!{\big(t\, \mu_{\rm I}E_{n_u}\!\!\!-\!\left|2n_u q_uB \right|\!-\!m_u\Delta m_l\big)h(0,n_u, n_d)\over 4E_{n_u}\left[-\left( tE_{n_u}\!-\,\mu_{\rm I}\right) ^2\!+\!E_{n_d}^2\right]}\tanh\!{E_{n_u}\!\!\!+\!t\,{\mu_{\rm I}\over2}\over 2T}\!+\!(u\!\leftrightarrow \!d)
\eea
with $\Delta m_l\equiv m_u-m_d$ and
\bea
\bar{h}(1,n_u, n_d)\equiv {32\pi\over |eB|^2}{({n_u}+{n_d}-1)!\over{n_u}!\,{n_d}!}
{\left({1\over3}\right)^{n_u-1}\left({2\over3}\right)^{n_d-1}}\theta({n_u}+{n_d}-1).
\eea
Utilizing the decomposition $\tanh{E_{n_u}+t\,{\mu_{\rm I}\over2}\over 2T}=1-{2/\left[\exp\left({E_{n_u}+t\,{\mu_{\rm I}\over2}\over T}\right)+1\right]}$, the divergent and convergent parts of the self-energies can be separated in the GL coefficients, that is, ${\cal A}_{\pi^\pm}={1\over2G_{11}^+}+\Pi^{0}_{\pi^\pm}+\Pi^{T}_{\pi^\pm}$ and ${\cal A}_{\rho^\pm_{\pm1}}={1\over2G_V}+\Pi^{0}_{\rho^\pm_{\pm1}}+\Pi^{T}_{\rho^\pm_{\pm1}}$ with the self-energy terms
\bea
\Pi^{0}_{\pi^\pm}&=&-{|q_uB|\over4\pi}{|q_dB|\over4\pi}\int{\di k_3\over 2\pi}\sum_{t=\pm}\sum_{n_u=0}^{\infty}\sum_{n_d=0}^{\infty}{\big( t\, \mu_{\rm I}E_{n_u}\!-\!m_u\Delta m_l\big)\left({\left| n_uq_uB \right|}\!+\!{\left| n_dq_dB \right|}\right)
\!-\!\left|2n_u q_uB \right|\left({\left| n_uq_uB \right|}\!-\!{\left| n_dq_dB \right|}\right)\over 4E_{n_u}\left[-\left( tE_{n_u}-\,\mu_{\rm I}\right) ^2+E_{n_d}^2\right]}\nonumber\\
&&\times{\bar{h}(1,n_u, n_d)}+(u\leftrightarrow d)\nonumber\\
&=&-{|q_uB|\over4\pi}{|q_dB|\over4\pi}\int{\di k_3\over 2\pi}\sum_{n_u=0}^{\infty}\sum_{n_d=0}^{\infty}{\bar{h}(1,n_u, n_d)}{E_{n_u}+E_{n_d}\over 4E_{n_u}E_{n_d}}\left[ {\left({\left| n_uq_uB \right|}\!+\!{\left| n_dq_dB \right|}\right)g(n_u, n_d)
}+{8{\left| n_uq_uB \right|}{\left| n_dq_dB \right|}
\over\left(E_{n_u}+E_{n_d}\right)^2}\right]\nonumber\\
&&\left[1+{\mu_I^2\over \left(E_{n_u}+E_{n_d}\right)^2-\mu_I ^2}\right] ,\\
\Pi^{T}_{\pi^\pm}&=&{|q_uB|\over4\pi}{|q_dB|\over4\pi}\int{\di k_3\over 2\pi}\sum_{t=\pm}\sum_{n_u=0}^{\infty}\sum_{n_d=0}^{\infty}{\big( t\, \mu_{\rm I}E_{n_u}\!-\!m_u\Delta m_l\big)\left({\left| n_uq_uB \right|}\!+\!{\left| n_dq_dB \right|}\right)
\!-\!\left|2n_u q_uB \right|\left({\left| n_uq_uB \right|}\!-\!{\left| n_dq_dB \right|}\right)\over 2E_{n_u}\left[-\left( tE_{n_u}-\,\mu_{\rm I}\right) ^2+E_{n_d}^2\right]}\nonumber\\
&&\times{\bar{h}(1,n_u, n_d)\over e^{E_{n_u}+t\,{\mu_{\rm I}\over2}\over T}+1}+(u\leftrightarrow d);\\
\nonumber\\
\nonumber\\
\Pi^{0}_{\rho^\pm_{\pm1}}&=&-{|q_uB|\over2\pi}{|q_dB|\over2\pi}\int{\di k_3\over 2\pi}\sum_{t=\pm}\sum_{n_u=0}^{\infty}\sum_{n_d=0}^{\infty}h(0,n_u, n_d){\big(t\, \mu_{\rm I}E_{n_u}\!-\!\left|2n_u q_uB \right|\!-\!m_u\Delta m_l\big)\over 4E_{n_u}\left[-\left( tE_{n_u}-\,\mu_{\rm I}\right) ^2+E_{n_d}^2\right]}+(u\leftrightarrow d)\nonumber\\
&=&-{|q_uB|\over2\pi}{|q_dB|\over2\pi}\int{\di k_3\over 2\pi}\sum_{n_u=0}^{\infty}\sum_{n_d=0}^{\infty}h(0,n_u, n_d)g(n_u, n_d){E_{n_u}+E_{n_d}\over 4E_{n_u}E_{n_d}}\left[{\mu_I^2\over \left(E_{n_u}+E_{n_d}\right)^2-\mu_I ^2}+1\right]\nonumber\\
\Pi^{T}_{\rho^\pm_{\pm1}}&=&{|q_uB|\over2\pi}{|q_dB|\over2\pi}\int{\di k_3\over 2\pi}\sum_{t=\pm}\sum_{n_u=0}^{\infty}\sum_{n_d=0}^{\infty}{\big(t\, \mu_{\rm I}E_{n_u}\!-\!\left|2n_u q_uB \right|\!-\!m_u\Delta m_l\big)\over 2E_{n_u}\left[-\left( tE_{n_u}-\,\mu_{\rm I}\right) ^2+E_{n_d}^2\right]}{h(0,n_u, n_d)\over e^{E_{n_u}+t\,{\mu_{\rm I}\over2}\over T}+1}+(u\leftrightarrow d).
\eea
Note that in the divergent parts $\Pi^{0}_{\pi^\pm}$ and $\Pi^{0}_{\rho^\pm_{\pm1}}$, we have defined an auxiliary function
\bea
g(n_u, n_d)\equiv{1}-{\left(\Delta m_l^2
+\left({2\left| n_uq_uB \right|}\!+\!{2\left| n_dq_dB \right|}\right)\right)
\over\left(E_{n_u}+E_{n_d}\right)^2},
\eea
and the vacuum and finite chemical potential terms are well separated in the last square brackets. 

The divergences in $\Pi^{0}_{\pi^\pm}$ and $\Pi^{0}_{\rho^\pm_{\pm1}}$ arise from the summations over the Landau levels and the energy-momentum of the internal quarks. We therefore regularize the quark part while keeping the mesonic part intact — that is, taking the limit $B\rightarrow0$ except for $P_n\left(\Delta{\bf x}_\bot\right)$ in \eqref{Dpi} and \eqref{Drho} for the counterterms. This scheme is more physical since we are renormalizing inverse propagators of charged mesons with given Landau levels in a given magnetic field. The renormalized terms vanish at zero magnetic field, thus we need to compensate regularized counterterms to reproduce the well-known results in the vanishing $B$ limit. According to Ref.~\cite{Ke:2026npb}, three finite terms are involved in each regularized $\Pi^{0}_{\pi^\pm/\rho^\pm_{\pm1}}$, that is, $\Pi^{0r}_{\pi^\pm/\rho^\pm_{\pm1}}=\Pi^{\Lambda}_{\pi^\pm/\rho^\pm_{\pm1}}+\Pi^{\rm B}_{\pi^\pm/\rho^\pm_{\pm1}}+\Pi^{\rm B,\mu_I}_{\pi^\pm/\rho^\pm_{\pm1}}$ with the explicit expressions as follows:
\bea
\Pi^{\Lambda}_{\pi^\pm}&=&-N_c\int_0^\Lambda\!\! {2k^2dk\over\pi^2}\frac{(E_{\rm u}E_{\rm d}\!+\!{m_{\rm u}}{m_{\rm d}}\!+\!k^2)(E_{\rm u}\!+\!E_{\rm d})}{E_{\rm u}E_{\rm d}[(E_{\rm u}\!+\!E_{\rm d})^2\!-\!\mu_I^2]},\\
\Pi^{\rm B}_{\pi^\pm}&=&-{N_c\over4\pi^2}\!\!\int\!{\di s\over s}\!\!\int_{-1}^1\!\! {\di u}~\Bigg\{{e^{-s\left(m_u^2u^+\!+m_d^2u^-\right)}\over 1\!+\!eBs\,R(B, s, u)}\!\!\Bigg[\left(m_um_d\!+\!{1\over s}\right)\!\!{1\!-\!\tanh{B_{\rm u}^s}^+\tanh{B_{\rm d}^s}^-\over {\tanh{B_{\rm u}^s}^+\over B_{\rm u}^s}+{\tanh{B_{\rm d}^s}^-\over B_{\rm d}^{s}}}+{(1\!-\!\tanh^2{B_{\rm u}^s}^+)(1\!-\!\tanh^2{B_{\rm d}^s}^-)\over s\left({\tanh{B_{\rm u}^s}^+\over B_{\rm u}^s}+{\tanh{B_{\rm d}^s}^-\over B_{\rm d}^{s}}\right)^2}\nonumber\\
&&{1\over 1+eBs\,R(B, s, u)}\Bigg]-{e^{-s\left(m_u^2u^+\!+m_d^2u^-\right)}\over 1+eBs{1-u^2\over4}}\left[m_um_d\!+\!{1\over s}\left(1+{1\over 1+eBs{1-u^2\over4}}\right)\right]\Bigg\},\\
\Pi^{\rm B,\mu_I}_{\pi^\pm}&=&-{\mu_I^2}\int{\di k_3\over 2\pi}\!\left[{|q_uB|\over4\pi}{|q_dB|\over4\pi}\sum_{n_u=0}^{N}\sum_{n_d=0}^{2N}\bar{h}(1,n_u, n_d){ \left(E_{n_u}\!+\!E_{n_d}\right)^2{\left({\left| n_uq_uB \right|}\!+\!{\left| n_dq_dB \right|}\right)g(n_u, n_d)
}\!+\!{8{\left| n_uq_uB \right|}{\left| n_dq_dB \right|}}\over 4E_{n_u}E_{n_d}\left(E_{n_u}+E_{n_d}\right)\left(\left(E_{n_u}+E_{n_d}\right)^2-\mu_I ^2\right)}\right.\nonumber\\
&&\!\!\!\!\!\!\!\!\!\!\!\!\!\!\!\!\!\left.-{|q_uB'|\over4\pi}\!{|q_dB'|\over4\pi}\!\!\sum_{n_u=0}^{N'}\!\sum_{n_d=0}^{2N'}\!\!{\left(E_{n_u}'\!\!\!+\!E_{n_d}'\right)^2\!\!g'(n_u, n_d)\!\left(
h'(0,n_u\!\!-\!1, n_d)\!+\!h'(0,n_u, n_d\!-\!1)\right)\!+\!{4h'(1,n_u\!\!-\!1, n_d\!-\!1)q_uB'q_dB'}\over 2E_{n_u}'E_{n_d}'\left(E_{n_u}'\!+\!E_{n_d}'\right)\left(\left(E_{n_u}'+E_{n_d}'\right)^2-\mu_I ^2\right)}\right]\!\!;\\
\nonumber\\
\nonumber\\
\Pi^{\Lambda}_{\rho^\pm_{\pm1}}&=&\!-\!N_c\!\int_0^\Lambda\!\! {2k^2dk\over\pi^2}\frac{(E_{\rm u}E_{\rm d}\!+\!{m_{\rm u}}{m_{\rm d}}\!+\!{1\over3}k^2)(E_{\rm u}\!\!+\!\!E_{\rm d})}{E_{\rm u}E_{\rm d}[(E_{\rm u}\!+\!E_{\rm d})^2\!-\!\mu_I^2]}\!-\!{N_c}\!\int_0^\Lambda\!\! {k^2dk\over\pi^2}\left\{{{q_{\rm u}B}\over(E_{\rm u}\!\!+\!\!E_{\rm d})^2}\left[\left(\frac{E_{\rm u}E_{\rm d}\!+\!{m_{\rm u}}{m_{\rm d}}\!+\!{1\over3}k^2}{E_{\rm u}^3}\!+\!{1\over E_{\rm u}}\!+\!{1\over E_{\rm d}}\right)\right.\right.\nonumber\\
&&\left.\left.-\frac{[(m_{\rm u}\!-\!m_{\rm d})^2\!+\!{4\over3}k^2]}{E_{\rm u}^2E_{\rm d}}\right]\!-\!(u\leftrightarrow d)\right\},\\
\Pi^{\rm B}_{\rho^\pm_{\pm1}}&=&\!-{N_c\over4\pi^2}\!\!\int\!{\di s\over s}\!\!\int_{-1}^1\!\! {\di u}~{e^{-s\left(m_u^2u^+\!\!+m_d^2u^-\right)}}\!\!\left(\!m_um_d\!+\!{1\over s}\right)\!\!\left[{\left(1\!+\!\tanh{B_{\rm u}^s}^+\right)\!\left(1\!-\!\tanh{B_{\rm d}^s}^-\right)\over [1\!+\!eBs\,R(B, s, u)]\left({\tanh{B_{\rm u}^s}^+\over B_{\rm u}^s}\!+\!{\tanh{B_{\rm d}^s}^-\over B_{\rm d}^{s}}\right)}\!-\!{1\!\!+\!{B_{\rm u}^s}^+\!\!\!-\!\!{B_{\rm d}^s}^-\over 1\!+\!{1-u^2\over 4}eBs}\right]\!\!,\\
\Pi^{\rm B,\mu_I}_{\rho^\pm_{\pm1}}&=&-\mu_I^2\int{\di k_3\over 2\pi}\!\Bigg[{|q_uB|\over2\pi}{|q_dB|\over2\pi}\sum_{n_u=0}^{N}\sum_{n_d=0}^{2N}\!{{h(0,n_u, n_d)}g(n_u, n_d, k_3)\left(E_{n_u}\!\!+\!\!E_{n_d}\right)\over 4\,E_{n_u}E_{n_d}\left(\left(E_{n_u}+E_{n_d}\right)^2-\mu_I ^2\right)}-\!{|q_uB'|\over2\pi}{|q_dB'|\over2\pi}\sum_{n_u=0}^{N'}\sum_{n_d=0}^{2N'}\!{{h'(0,n_u, n_d)}\over 4\,E_{n_u}'E_{n_d}'}\nonumber\\
&&{g'(n_u, n_d, k_3)(E_{n_u}'\!+\!E_{n_d}')\over \left(E_{n_u}'+E_{n_d}'\right)^2-\mu_I ^2}\!\Bigg].
\eea
Here, $B_{f}^{s}\equiv q_fBs, {B_{\rm u}^s}^+\equiv q_uBs{1+u\over2},  {B_{\rm d}^s}^+\equiv q_dBs{1-u\over2}$, and the involved auxiliary functions are defined as $g'(n_u, n_d, k_3)\equiv g(n_u, n_d, k_3)|_{B\rightarrow B'}$, $R(B, s, u)\equiv{{\tanh{B_{\rm u}^s}^+\over B_{\rm u}^s}{\tanh{B_{\rm d}^s}^-\over B_{\rm d}^{s}}}/\left({{\tanh{B_{\rm u}^s}^+\over B_{\rm u}^s}+{\tanh{B_{\rm d}^s}^-\over B_{\rm d}^{s}}}\right)$ from the convolution with $P_0\left({\bf q}_\bot^2\right)$ according to \eqref{DM}, and
\bea
\!\!\!\!\!\!h'(\alpha, n_u, n_d)&=&{16\pi\over |eB|}\left({2\over|eB|}\right)^{\alpha}\frac{(n_u\!+\!n_d\!+\!\alpha)!}{n_u!\,n_d!}
\frac{(b'\!-\!\tilde{q}_u')^{n_u}(b'\!-\!\tilde{q}_d')^{n_d}}{{b'}^{n_u+n_d+\alpha+1}}\,
{}_2F_1\!\left(
\!-n_u,-n_d;\,-n_u\!-\!n_d\!-\!\alpha;\,
\frac{b'(b'\!-\!\tilde{q}_u'\!-\!\tilde{q}_u')}{(b'\!-\!\tilde{q}_u')(b'\!-\!\tilde{q}_d')}
\right)
\eea
with $b'\equiv{1+\tilde{q}_u'+\tilde{q}_d'\over2}, \tilde{q}_u'\equiv q_u B'/(eB)$, and $\tilde{q}_d'\equiv q_d B'/(eB)$. Note that the counterterms in $\Pi^{\rm B}_{\pi^\pm/\rho^\pm_{\pm1}}$ and $\Pi^{\rm B,\mu_I}_{\pi^\pm/\rho^\pm_{\pm1}}$ are $B$-dependent, following the expressions of $P_n\left({\bf \Delta x}_\bot\right)$.

In total, the regularized Ginzburg-Landau coefficients are ${\cal A}_{\pi^\pm}={1\over2G_{11}^+}+\Pi^{\Lambda}_{\pi^\pm}+\Pi^{\rm B}_{\pi^\pm}+\Pi^{\rm B,\mu_I}_{\pi^\pm}+\Pi^{T}_{\pi^\pm}$ for ${\pi^\pm}$ and ${\cal A}_{\rho^\pm_{\pm1}}={1\over2G_V}+\Pi^{\Lambda}_{\rho^\pm_{\pm1}}+\Pi^{\rm B}_{\rho^\pm_{\pm1}}+\Pi^{\rm B,\mu_I}_{\rho^\pm_{\pm1}}+\Pi^{T}_{\rho^\pm_{\pm1}}$ for ${\rho^\pm_{\pm1}}$, respectively.
\end{widetext}

\section{Numerical results}\label{num}
\begin{figure}[!htb]
   \begin{center}
        \includegraphics[width=8cm]{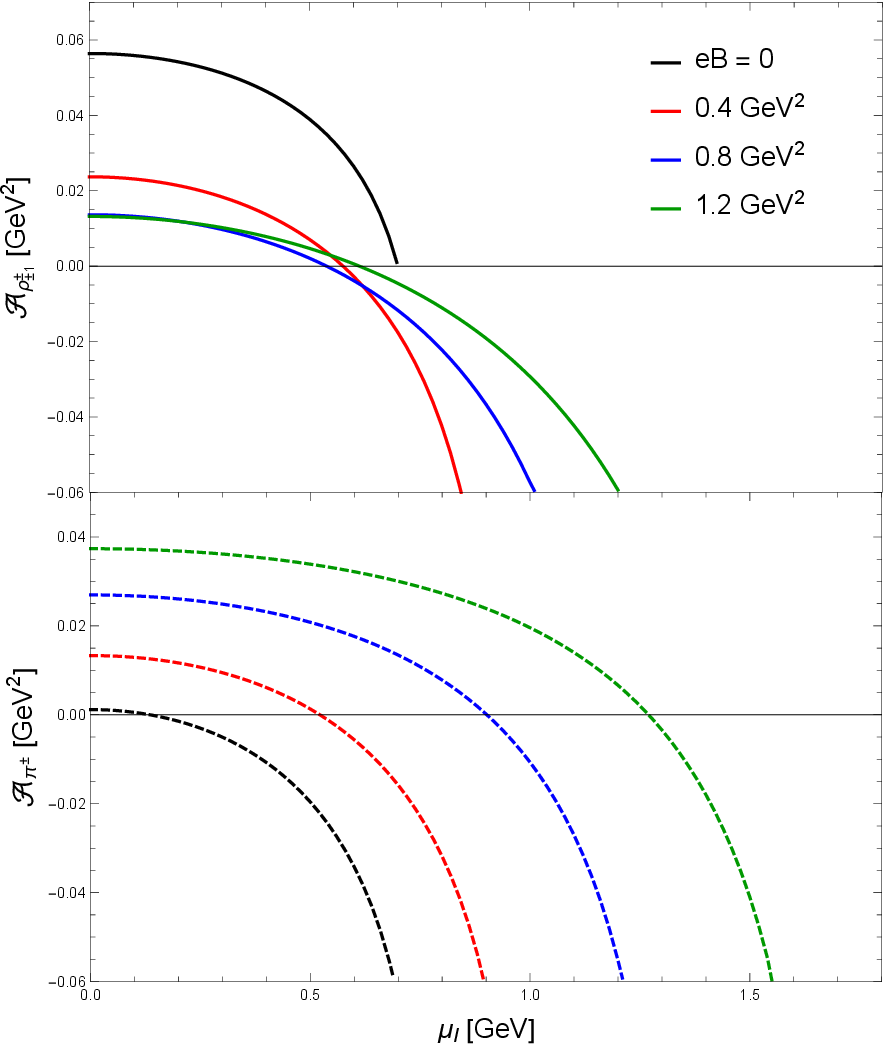}
      \caption{The Ginzburg–Landau coefficients $\mathcal{A}_{\pi^\pm}$ (dashed lines) and $\mathcal{A}_{{\rho}_{\pm1}^\pm}$ (solid lines) as functions of the isospin chemical potential $\mu_I$ for four different magnetic fields, ${eB}=0$ (black), $0.4$ (red), $0.8$ (blue), and $1.2\,\mathrm{GeV}$ (green). Note that the points where $\mathcal{A}=0$ correspond to the onsets of second-order transitions and thus determine the phase boundary.}
    \label{GLs}
      \end{center}
\end{figure}

In order to carry out numerical calculations, we choose the following parameters for the scalar-pseudoscalar sector: $m_{\rm u}=m_{\rm d}=5.5~{\rm MeV}, m_{\rm s}=140.7~{\rm MeV}, \Lambda=602.3~{\rm MeV}, G_S\Lambda^2=1.835,$ and $K\Lambda^5=12.36$~\cite{Rehberg:1995kh}. To avoid artifacts, the vector coupling constant is fixed to $G_V\Lambda^2=1.522$ by fitting to a vacuum mass of the $\rho$ meson, $m_\rho^v=0.7~{\rm GeV}$, smaller than the true value~\cite{Cao:2019res}. The Ginzburg–Landau coefficients $\mathcal{A}$ for ${\pi^\pm}$ and ${{\rho}_{\pm1}^\pm}$ are demonstrated together in Fig.~\ref{GLs} for four different magnetic fields. As we can see, while the coefficient $\mathcal{A}_{\pi^\pm}$ increases with $eB$ for a given $\mu_I$, the coefficient $\mathcal{A}_{{\rho}_{\pm1}^\pm}$ shows a non-monotonic behavior with $eB$. The latter follows the feature of the lowest energy of ${{\rho}_{\pm1}^\pm}$ found previously in the three-flavor NJL model~\cite{Cao:2019res}, but the true feature could be that $\mathcal{A}_{{\rho}_{\pm1}^\pm}$ decreases monotonically with $eB$ and then saturates according to lattice QCD simulations~\cite{Bali:2017ian}.

\begin{figure}[!htb]
	\begin{center}
	\includegraphics[width=8cm]{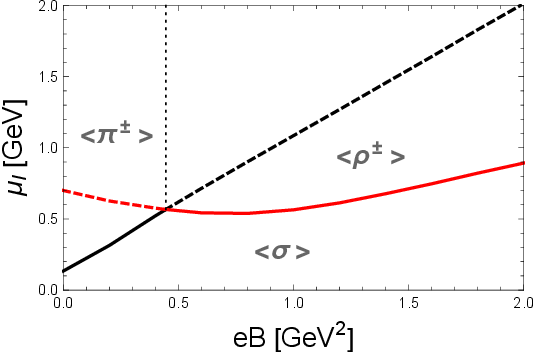}
	\caption{The phase transition lines for pion superfluidity (black) and rho superconductivity (red) in the $eB$-$\mu_I$ plane with the solid ones favored. The notations $\langle\sigma\rangle, \langle\pi^\pm\rangle$, and $\langle\rho^\pm\rangle$ correspond to the normal chiral symmetry breaking phase, pion superfluidity, and rho superconductivity, respectively. The thin dotted line sketches a possible boundary between pion superfluidity and rho superconductivity.}\label{PT}
	\end{center}
\end{figure}

The points where $\mathcal{A}=0$ correspond to the onsets of second-order transitions and thus are typically illustrated in Fig.~\ref{PT}. The results are qualitatively consistent with the two-flavor case~\cite{Ke:2026npb}: as the isospin chemical potential increases, pion superfluidity is favored at small magnetic fields, while rho superconductivity is favored at large magnetic fields. However, the phase boundary between the normal chiral symmetry breaking phase and rho superconductivity increases with larger magnetic field, still a demonstration of the feature of the lowest energy of ${{\rho}_{\pm1}^\pm}$ in the three-flavor NJL model~\cite{Cao:2019res}.

\section{Summary}\label{summary}
In this work, we re-explore the QCD phase transition at finite isospin density and magnetic field within the extended three-flavor Nambu--Jona-Lasinio model by considering both pion superfluidity and rho superconductivity. According to the previous discussions in the two-flavor case~\cite{Ke:2026npb}, both phase transitions are of second order, thus we adopt the Ginzburg-Landau approximation and the Landau representation for fermion propagators to pin down the phase boundaries. In our previous calculations of self-energies~\cite{Ke:2026npb}, the overall Schwinger phases from quark loops were dropped by hand to keep the results gauge invariant. In this work, we more physically project the mesonic fields onto the eigenstates of a charged point particle in a magnetic field, where gauge invariance is self-consistently guaranteed for the self-energies. Moreover, the self-energies are proven to be degenerate with respect to the extra transverse degrees of freedom, such as the longitudinal angular momentum $l$ in the symmetric gauge. Nevertheless, the new numerical results are qualitatively consistent with previous findings~\cite{Ke:2026npb}: as the isospin chemical potential increases, pion superfluidity is favored at small magnetic fields, while rho superconductivity is favored at large magnetic fields. According to our previous study in the three-flavor NJL model~\cite{Cao:2019res}, the lowest energy of the rho meson increases with stronger magnetic field due to the mass splitting between $u$ and $d$ quarks, inconsistent with lattice QCD simulations~\cite{Bali:2017ian}. It is therefore natural that the corresponding critical isospin chemical potential also enhances with stronger magnetic field here.

As mentioned in our previous work~\cite{Ke:2026npb}, it is challenging to explore the transition between pion superfluidity and rho superconductivity as the magnetic field increases at a large isospin chemical potential. Though the magnetic field can introduce mixing between $\pi^\pm$ and the longitudinal-spin-zero mode $\rho^\pm_3$~\cite{Bali:2017ian}, such a consideration will not significantly affect our exploration of the competition between $\pi^\pm$ and $\rho^\pm_{\pm 1}$ condensations. Based on this work, the nontrivial QCD phase transitions in the early Universe~\cite{Vovchenko:2020crk,Middeldorf-Wygas:2020glx,Cao:2021gfk,Cao:2022fow,Cao:2024fyk} can be more realistically explored.

\section*{Acknowledgment}
G.C. is funded by the National Natural Science Foundation of China with Grant Nos. 12447102 and 12575152, and the Natural Science Foundation of Guangdong Province with Grant No. 2024A1515011225.

\appendix
\begin{widetext}
\section{Proof of the degeneracy with respect to the longitudinal angular momentum}\label{degen}

For a meson composed of $u$ and $\bar{d}$ quarks, that is, $M=\bar{d}\,\Gamma_{\rm M}\,u$, the polarization function can be evaluated as 
\bea
\Pi_{n,l}(q_3,q_4)\equiv-{1\over V_4}\int{\di^4x}{\di^4x'}{e^{-i[q_3(x_3-x_3')+q_4(x_4-x_4')]}}\chi_n^{l*}(eB,x_\bot)\chi_n^{l}(eB,x_\bot'){\rm Tr}\ G_{u}(x,x')\Gamma_{\rm M}^\dagger G_{d}(x',x)\Gamma_{\rm M}\label{Pin}
\eea
according to \eqref{Dpi0} and \eqref{Drho0}. For a point-like charged meson in the magnetic field, we know that the eigenenergies are degenerate with respect to $l$. In the following, we are going to prove that this degeneracy remains valid even when the polarization loop is considered. 

As $l$ is only relevant to the transverse dynamics, we will just focus on the transversal coordinates, that is, ${\bf x}_\bot$ and ${\bf x}_\bot'$. After completing the trace, the relevant part becomes formally
\bea
\Pi_{n,l}^\bot\equiv-\int{\di^2x_\bot}{\di^2x_\bot'}\chi_n^{l*}(eB,x_\bot)\chi_n^{l}(eB,x_\bot')e^{ i\Phi(eB,x_\bot,x'_\bot)-{\left| eB \right|\over4} {\bf \Delta x}_\bot^2} \sum_{n_u,n_d=0}^{\infty}f_{n_d,q_d}^{n_u,q_u}({\Delta x}_1,{\Delta x}_2),
\eea
where $f_{n_d,q_d}^{n_u,q_u}({\Delta x}_1,{\Delta x}_2)$ are polynomial functions of ${\Delta x}_\bot$ with the factors depending on $n_f$ and $q_f\ (f=u,d)$.  The reason why the transversal part must be in this form is that: apart from the exponential part, the generalized Laguerre polynomials $L_n^l(x)$ are themselves polynomials of $x$ with the highest order $x^n$. To guarantee that $l$ is a good quantum number, the index $\Gamma_{\rm M}$ must be chosen such that $f_{n_d,q_d}^{n_u,q_u}({\Delta x}_1,{\Delta x}_2)$ depends compactly on a single variable, that is, ${\bf \Delta x}_\bot^2$, such as the case for $\rho^\pm$. Then, we can rearrange the functions as
\bea
f_{n_d,q_d}^{n_u,q_u}({\bf \Delta x}_\bot^2)=\sum_{n=0}^{n_u+n_d}C(n_u,q_u;n_d,q_d;n,e)\, L_n\left({|eB|\over2}{\bf \Delta x}_\bot^2\right),
\eea
where the expansion factor $C(n_u,q_u;n_d,q_d;n,e)$ is $l$-independent, and $L_n\left({|eB|\over2}{\bf \Delta x}_\bot^2\right)$ is the Laguerre polynomial with the highest order term $\left({|eB|\over2}{\bf \Delta x}_\bot^2\right)^n$ and satisfies the expansion
\bea
{|eB|\over2\pi}e^{ i\Phi(eB,x_\bot,x'_\bot)-{\left| eB \right|\over4} {\bf \Delta x}_\bot^2} L_n\left({|eB|\over2}{\bf \Delta x}_\bot^2\right)=\sum_{l=-n}^\infty\chi_n^{l*}(eB,x_\bot')\chi_n^{l}(eB,x_\bot).
\eea
Note that the commonly involved function $e^{ i\Phi(eB,x_\bot,x'_\bot)-{\left| eB \right|\over4} {\bf \Delta x}_\bot^2}$ is important to guarantee that the expansions correspond to a meson with a charge $e=q_u+q_{\bar d}=|q_u|+|q_d|$. 

Eventually, the transversal part of the polarization function can be evaluated as
\bea
\Pi_{n,l}^\bot &\equiv&{-2\pi\over|eB|}\!\sum_{n_u,n_d=0}^{\infty}\!\!\sum_{n'=0}^{n_u\!\!+\!n_d}\!\!\!C(n_u,q_u;n_d,q_d;n',e)\!\!\!\!\sum_{l'=-n'}^\infty\!\int\!{\di^2x_\bot}{\di^2x_\bot'}\chi_n^{l*}(eB,x_\bot)\chi_n^{l}(eB,x_\bot')\chi_{n'}^{l'*}(eB,x_\bot')\chi_{n'}^{l'}(eB,x_\bot)\nonumber\\
&=&{-2\pi\over|eB|}\!\sum_{n_u,n_d=0}^{\infty}\!\sum_{n'=0}^{n_u+n_d}\!\!C(n_u,q_u;n_d,q_d;n',e)\sum_{l'=-n'}^\infty\!\!\delta_{n,n'}\delta_{l,l'}={2\pi\over|eB|}\!\sum_{n_u,n_d=0}^{\infty}\!\!C(n_u,q_u;n_d,q_d;n,e)\Theta(n_u\!+\!n_d\!-\!n)\nonumber\\
\eea
due to the orthogonality and normalization of $\chi_n^{l}(eB,x_\bot)$. As we can see, the polarization function is $l$-independent.

\section{The explicit forms of the inverse propagators}\label{prop2}

As we do not know the explicit form of the dimensionless factor $C(n_u,q_u;n_d,q_d;n,e)$, it is more convenient to sum over $l$ in Eq.~\eqref{Pin} in order to evaluate the polarization function explicitly. Then, a Schwinger phase $e^{ i\Phi(eB,x_\bot',x_\bot)}$ shows up to exactly cancel the one from quark loops. The left term is translationally invariant, thus it can be transformed to energy-momentum space~\cite{Miransky:2015ava}. As the quark propagators can be formally expressed as $G_{f}(x,x')\equiv e^{ i\Phi(q_fB,x_\bot,x_\bot')}\tilde{G}_{f}(\Delta x)$, the $l$-summed inverse propagator of the meson can be evaluated as
\bea
D_{\rm M}^{-1}(q_\parallel,n)&=&\left({|eB|\over 2\pi}V_4\right)^{-1}\Bigg[\int \di^4x\sum_{l=-n}^\infty{\chi_n^{l*}(eB,x_\bot)\chi_n^{l}(eB,x_\bot)\over4G}-\int{\di^4x}{\di^4x'}\sum_{l=-n}^\infty{e^{-i(q_3\Delta x_3+q_4\Delta x_4)}}\nonumber\\
&&\ \ \ \ \ \ \ \ \ \ \ \ \ \ \ \ \ \ \ \ \ \times\chi_n^{l*}(eB,x_\bot)\chi_n^{l}(eB,x_\bot')e^{ i\Phi(eB,x_\bot,x'_\bot)}{\rm Tr}\ \tilde{G}_{u}(\Delta x)\Gamma_{\rm M}^\dagger \tilde{G}_{d}(-\Delta x)\Gamma_{\rm M}\Bigg]\nonumber\\
&=&{1\over4G_{\rm M}}-\int{\di^4\Delta x}\ {e^{-i(q_3\Delta x_3+q_4\Delta x_4)}}P_{\rm n}({\bf \Delta x}_\bot)\ {\rm Tr}\ \tilde{G}_{u}(\Delta x)\Gamma_{\rm M}^\dagger \tilde{G}_{d}(-\Delta x)\Gamma_{\rm M}
\eea
with the transversal projecting function $P_{n}({\bf \Delta x}_\bot)\equiv e^{-{\left| eB \right|\over4} {\bf \Delta x}_\bot^2} L_n\left({|eB|\over2}{\bf \Delta x}_\bot^2\right)=P_{n}(-{\bf \Delta x}_\bot)$. 

As we are interested in the dynamical features, it is usually more convenient to shift to the energy-momentum space and we have
\bea
D_{\rm M}^{-1}(q_\parallel,n)={1\over4G_{\rm M}}-\int{\di^4k\over(2\pi)^4}\int{\di^2{\bf q_\bot}\over(2\pi)^2}\ P_n\left(-{\bf q_\bot}\right)\ {\rm Tr}\ \tilde{G}_{u}(k+q)\Gamma_{\rm M}^\dagger \tilde{G}_{d}(k)\Gamma_{\rm M}\label{DM}
\eea
where $\tilde{G}_{f}(k)$ and $P_n\left({\bf q_\bot}\right)$ are Fourier transforms of $\tilde{G}_{f}(\Delta x)$ and $P_n({\bf \Delta x}_\bot)$, respectively, that is,
\bea
\tilde{G}_{f}(k)=\int\di^4x\ e^{i\, k\cdot \Delta x}\tilde{G}_{f}(\Delta x),\ \ P_n\left({\bf q_\bot}\right)=\int\di^2x_\bot e^{-i\, {\bf q_\bot\cdot \Delta x_\bot}} P_n({\bf \Delta x}_\bot)=P_n\left(-{\bf q_\bot}\right).
\eea
According to Ref.~\cite{Miransky:2015ava}, $P_n\left({\bf q_\bot}\right)=(-1)^n{4\pi\over|eB|}e^{-{{\bf q}_\bot^2\over\left|eB \right|}}L_n\left({2{\bf q}_\bot^2\over\left|eB \right|}\right)$, so we are able to evaluate the dynamical features of the meson for any given Landau level $n$. Since ${|eB|\over2\pi}\sum_{n=0}^\infty P_n\left({\bf q_\bot}\right)=1$, the summation over the Landau level $n$ in \eqref{DM} gives a result as if the mesons and quarks are well defined in effective transverse momentum space. Of course, only when $eB\rightarrow 0$ are the transverse momenta truly well-defined conserved quantities. Notice that $P_0\left({\bf q_\bot}\right)={4\pi\over|q_MB|}e^{-{{\bf q}_\bot^2\over\left| q_MB \right|}}$ here, compared to $P_0=(2\pi)^2\delta_{{\bf q}_\bot,0}$ in our previous inconsistent treatment~\cite{Cao:2015xja}. However, in the limit $q_MB\rightarrow0$, ${4\pi\over|q_MB|}e^{-{{\bf q}_\bot^2\over\left| q_MB \right|}}\rightarrow (2\pi)^2\delta_{{\bf q}_\bot,0}$, so the previous treatment actually corresponds to the small magnetic field approximation.

In \eqref{DM}, ${\bf q_\bot}$ is not the true transverse momentum of the meson and the integral dimension increases in the full energy momentum space. To avoid complication, it is more feasible to work in the mixed spaces, that is, in the energy-momentum space for the longitudinal dynamics but in the coordinate space for the transverse dynamics. Then, the inverse meson propagator can be rewritten as
\bea
D_{\rm M}^{-1}(q_\parallel,n)={1\over4G_{\rm M}}-\int{\di^2k_\parallel\over(2\pi)^2}\int{\di^2{\bf \Delta x_\bot}}\ P_n\left({\bf \Delta x_\bot}\right)\ {\rm Tr}\ \tilde{G}_{u}(k_\parallel+q_\parallel, {\bf \Delta x_\bot})\Gamma_{\rm M}^\dagger \tilde{G}_{d}(k_\parallel,-{\bf \Delta x_\bot})\Gamma_{\rm M}
\eea
with $\tilde{G}_{f}(k_\parallel+q_\parallel, {\bf \Delta x_\bot})$ following \eqref{Sfx} as
\bea
\tilde{G}_{f}(k_\parallel+q_\parallel, {\bf \Delta x_\bot})&=&-{i|q_fB|\over4\pi}e^{-{\left| q_fB \right|\over4} {\bf \Delta x}_\bot^2} \sum_{n=0}^{\infty}{\frac{D_n(q_fB,{\bf \Delta x}_\bot)}{\left( k_{4}^{f} \right) ^2+k_{3}^{2}+m_f^2+2\left| q_fB \right|n}}.
\eea
More explicitly, the inverse meson propagator is
\bea
&&D_{\rm M}^{-1}(q_\parallel,n)={1\over4G}+{|q_uB|\over4\pi}{|q_dB|\over4\pi}\int{\di^2k_\parallel\over(2\pi)^2}\sum_{n_u=0}^{\infty}\sum_{n_d=0}^{\infty}{N(n, n_u, n_d; eB, q_uB, q_dB)\over \left[\left( k_{4}^{u} +q_4\right) ^2+E_{n_u}^2(k_3+q_3)\right]\left[\left( k_{4}^{d} \right) ^2+E_{n_d}^2\right]},\\
&&N(n, n_u, n_d; eB, q_uB, q_dB)\equiv \int{\di^2{\bf \Delta x_\bot}}e^{-{\left| eB \right|\over2} {\bf \Delta x}_\bot^2} L_n\left({|eB|\over2}{\bf \Delta x}_\bot^2\right){\rm Tr}\ D_{n_u}(q_uB,{\bf \Delta x}_\bot)\Gamma_{\rm M}^\dagger D_{n_d}(q_dB,-{\bf \Delta x}_\bot)\Gamma_{\rm M}\nonumber
\eea
with $E_{n_f}(k_3)\equiv\sqrt{k_3^2+2n_f|q_fB|+m_f^2}$.

For $\pi^+$ and $\bar{\rho}_1^{+}$, the interacting matrices are $\Gamma_{\pi^+}=i\gamma^5$ and $\Gamma_{\bar{\rho}_1^{+}}={\gamma_1+ i\gamma_2\over \sqrt{2}}$, respectively. So it follows that
\bea
N_{\pi^+}&=&
-\big[(k_3+q_3)k_3+k_4^{u}k_4^{d}+m_um_d\big]
\left[
h(0,n, n_u-1, n_d)+h(0,n, n_u, n_d-1)\right]\nonumber\\
&&
-2\left| q_uB \right|\left| q_dB \right|h(1,n, n_u-1, n_d-1)
\\
N_{\bar{\rho}_1^{+}}&=&-2\big[(k_3+q_3)k_3+k_4^{u}k_4^{d}+m_um_d\big]h(0,n, n_u, n_d)
\eea
with 
\bea
h(\alpha,n, n_u, n_d)&\equiv& 8\int{\di^2{\bf \Delta x_\bot}}e^{-{\left| eB \right|\over2} {\bf \Delta x}_\bot^2} {\bf \Delta x}_\bot^{2\alpha}L_n\left({|eB|\over2}{\bf \Delta x}_\bot^2\right)L_{n_u}^\alpha\!\left({\left| q_uB \right|{\bf \Delta x}_\bot^2\over2}\right)
L_{n_d}^\alpha\!\left({\left| q_dB \right|{\bf \Delta x}_\bot^2\over2}\right)\nonumber\\
&=&{16\pi\over |eB|}\left({2\over|eB|}\right)^{\alpha}\int_0^\infty\di{r}\ e^{-r } r^{\alpha}L_n\left(r\right)L_{n_u}^\alpha\!\left({2\over3}r\right)
L_{n_d}^\alpha\!\left({1\over3}r\right).
\eea

\subsection*{Examples for the Landau levels $n=0-3$}\label{exmp}
By using the expansion relation $L_n\left(r\right)=\sum_{p=0}^n{(-1)^pn!r^p\over(n-p)!(p!)^2}$, the recurrence relation~\cite{Abramowitz1972}  
\bea
&&r\,L_n^\alpha\left(r\right)=(2n+\alpha+1)L_n^\alpha\left(r\right)-(n+1)L_{n+1}^\alpha\left(r\right)-(n+\alpha)L_{n-1}^\alpha\left(r\right)=\sum_{p=0}^2{2(-1)^{p+1}\over p!(2-p)!}\left[(n+1-p)+p{\alpha+1\over2}\right]L_{n+1-p}^\alpha\left(r\right),
\eea
and the standard integral identities for generalized Laguerre polynomials~\cite{Gradshteyn1980}, 
\bea
\int_{0}^{\infty}\!\!\!dx\,e^{-bx}\,x^{\alpha}\,L_{n}^{\alpha}(\lambda x)\,L_{p}^{\alpha}(\mu x)
=\frac{\Gamma(p\!+\!n\!+\!\alpha\!+\!1)}{p!\,n!}
\frac{(b\!-\!\lambda)^n(b\!-\!\mu)^p}{b^{p+n+\alpha+1}}\,
{}_2F_1\!\left(
-p,-n;\,-p-n-\alpha;\,
\frac{b(b\!-\!\lambda\!-\!\mu)}{(b\!-\!\lambda)(b\!-\!\mu)}
\right),
\eea
the function $h(\alpha,n, n_u, n_d)$ can be evaluated explicitly, see the following examples for the Landau levels $n=0-3$.\\
\\
${\bf (a)\ n=0}$, $L_0\left(r\right)=1$ and
\bea
h(\alpha,0, n_u, n_d)
&=&{16\pi\over |eB|}\left({2\over|eB|}\right)^{\alpha}{({n_u}\!+\!{n_d}\!+\!\alpha)!\over{n_u}!\,{n_d}!}
{\left({1\over3}\right)^{n_u}\left({2\over3}\right)^{n_d}}.
\eea
\\
${\bf (b)\ n=1}$, $L_1\left(r\right)=1-x$ and
\bea
h(\alpha,1, n_u, n_d)&=&h(\alpha,0, n_u, n_d)-{16\pi\over |eB|}\left({2\over|eB|}\right)^{\alpha}\!\int_0^\infty\!\di{r}\ e^{-r }\! r^{\alpha}L_{n_u}^\alpha\!\left({2\over3}r\right)\left[(2{n_d}\!+\!\alpha\!+\!1)L_{n_d}^\alpha\left({1\over3}r\right)\!-\!({n_d}\!+\!1)L_{{n_d}\!+\!1}^\alpha\left({1\over3}r\right)\right.\nonumber\\
&&\left.-({n_d}+\alpha)L_{{n_d}-1}^\alpha\left({1\over3}r\right)\right]\\
&=&h(\alpha,0, n_u, n_d)\!-\!{1\over|\tilde{q}_{d}|}{16\pi\over |eB|}\!\left({2\over|eB|}\right)^{\alpha}\!{({n_u}\!\!+\!{n_d}\!+\!\alpha)!\over{n_u}!\,{n_d}!}
{\left({1\over3}\right)^{n_u}\!\!\left({2\over3}\right)^{n_d}}\!\!\left[(2{n_d}\!+\!\alpha\!+\!1)\!-\!(1\!-\!|\tilde{q}_{d}|){({n_u}\!\!+\!{n_d}\!+\!\alpha\!+\!1)}\right.\nonumber\\
&&\left.-{1\over 1-|\tilde{q}_{d}|}{n_d({n_d}+\alpha)\over {n_u}\!+\!{n_d}\!+\!\alpha}\right]\nonumber\\
&=&h(\alpha,0, n_u, n_d)-{16\pi\over |eB|}\left({2\over|eB|}\right)^{\alpha}{({n_u}\!+\!{n_d}\!+\!\alpha)!\over{n_u}!\,{n_d}!}
{\left({1\over3}\right)^{n_u}\left({2\over3}\right)^{n_d}}\left[-{{n_d}\over1-|\tilde{q}_{d}|}-{n_u\over|\tilde{q}_{d}|}+{({n_u}\!+\!{n_d}\!+\!\alpha+1)}\right.\nonumber\\
&&\left.+{1\over |\tilde{q}_{d}|(1-|\tilde{q}_{d}|)}{n_dn_u\over {n_u}\!+\!{n_d}\!+\!\alpha}\right]\nonumber\\
&=&h(\alpha,0, n_u, n_d)\!-\!{16\pi\over |eB|}\left({2\over|eB|}\right)^{\alpha}\!\!\!\sum_{p_u,p_d=0}^1\!(-1)^{p_u+p_d}{({n_u}\!-\!p_u\!+\!{n_d}\!-\!p_d\!+\!\alpha+1)!\over({n_u}-p_u)!\,({n_d}-p_d)!}
{\left({1\over3}\right)^{n_u-p_u}\left({2\over3}\right)^{n_d-p_d}}.
\eea
\\
${\bf (c)\ n=2}$, $L_2\left(r\right)=1-2x+x^2/2=2L_1\left(r\right)-L_0\left(r\right)+x^2/2$ and
\bea
h(\alpha,2, n_u, n_d)&=&2h(\alpha,1, n_u, n_d)-h(\alpha,0, n_u, n_d)+{36\pi\over |eB|}\left({2\over|eB|}\right)^{\alpha}\sum_{p_u,p_d=0}^2{2(-1)^{p_u+1}\over p_u!(2-p_u)!}\left[(n_u+1-p_u)+p_u{\alpha+1\over2}\right]\nonumber\\
&&{2(-1)^{p_d+1}\over p_d!(2-p_d)!}\left[(n_d+1-p_d)+p_d{\alpha+1\over2}\right]\int_0^\infty\di{r}\ e^{-r } r^{\alpha}L_{n_u+1-p_u}^\alpha\!\left({2\over3}r\right)L_{n_d+1-p_d}^\alpha\left({1\over 3}r\right)\nonumber\\
&=&2h(\alpha,1, n_u, n_d)-h(\alpha,0, n_u, n_d)+{8\pi\over |eB|}\left({2\over|eB|}\right)^{\alpha}\sum_{p_u,p_d=0}^2{4(-1)^{p_u+p_d}{\left({1\over3}\right)^{n_u-p_u}\left({2\over3}\right)^{n_d-p_d}}\over p_u!(2-p_u)!p_d!(2-p_d)!}\nonumber\\
&&{({n_u}\!-\!p_u\!+\!{n_d}-p_d\!+\!\alpha+2)!\over({n_u}\!+\!1\!-\!p_u)!\,({n_d}\!+\!1\!-\!p_d)!}\left[(n_u\!+\!1\!-\!p_u)\!+\!p_u{\alpha\!+\!1\over2}\right]\left[(n_d\!+\!1\!-\!p_d)\!+\!p_d{\alpha\!+\!1\over2}\right].
\eea
\\
${\bf (d)\ n=3}$, $L_3\left(r\right)=1-2x+x^2/2=3L_2\left(r\right)-3L_1\left(r\right)+L_0\left(r\right)-x^3/6$ and
\bea
&&h(\alpha,3, n_u, n_d)\nonumber\\
&=&3h(\alpha,2, n_u, n_d)-3h(\alpha,1, n_u, n_d)+h(\alpha,0, n_u, n_d)-{12\pi\over |eB|}\left({2\over|eB|}\right)^{\alpha}\sum_{p_u,p_d=0}^2{2(-1)^{p_u+1}\over p_u!(2-p_u)!}\left[(n_u\!+\!1\!-\!p_u)\!+\!p_u{\alpha\!+\!1\over2}\right]\nonumber\\
&&{2(-1)^{p_d+1}\over p_d!(2-p_d)!}\left[(n_d+1-p_d)+p_d{\alpha+1\over2}\right]\int_0^\infty\di{r}\ e^{-r } r^{\alpha+1}L_{n_u+1-p_u}^\alpha\!\left({2\over3}r\right)L_{n_d+1-p_d}^\alpha\left({1\over 3}r\right)\nonumber\\
&=&3h(\alpha,2, n_u, n_d)-3h(\alpha,1, n_u, n_d)+h(\alpha,0, n_u, n_d)-{8\pi\over 3|eB|}\left({2\over|eB|}\right)^{\alpha}\sum_{p_u,p_d=0}^2\sum_{p_u',p_d'=0}^1{4(-1)^{P_u\!+\!P_d}{\left({1\over3}\right)^{n_u\!-\!P_u}\left({2\over3}\right)^{n_d\!-\!P_d}}\over p_u!(2-p_u)!p_d!(2-p_d)!}\nonumber\\
&&\left[(n_u\!+\!1\!-\!p_u)\!+\!p_u{\alpha\!+\!1\over2}\right]\left[(n_d\!+\!1\!-\!p_d)\!+\!p_d{\alpha\!+\!1\over2}\right]{({n_u}\!-\!P_u\!+\!{n_d}\!-\!P_d\!+\!\alpha+3)!\over({n_u}+1-P_u)!({n_d}+1-P_d)!}
\eea
with $P_{f}=p_f+p_f'$.
\end{widetext}


\end{document}